\documentclass[superscriptaddress,onecolumn,secnumarabic,
amssymb,amsmath,nobibnotes,aps,prd,showkeys,showpacs,nofootinbib]{revtex4}

\usepackage[latin1]{inputenc}
\usepackage{graphicx}
\usepackage[english]{babel}

\usepackage{amsmath}
\usepackage{amssymb}
\usepackage{amsfonts}
\usepackage{colordvi}
\usepackage{psfrag}
\usepackage{color}

\begin{document}

\newcommand\cL{\mathcal{L}}
\newcommand\be{\begin{equation}}
\newcommand\ee{\end{equation}}
\newcommand\bea{\begin{eqnarray}}
\newcommand\eea{\end{eqnarray}}
\newcommand\beq{\begin{eqnarray}}
\newcommand\eeq{\end{eqnarray}}
\newcommand\tr{{\rm tr}\, }
\newcommand\nn{\nonumber \\}
\newcommand\e{{\rm e}}


\newcommand\bef{\begin{figure}}
\newcommand\eef{\end{figure}}
\newcommand{\ans}{ansatz }
\newcommand{\eeqn}{\end{eqnarray}}
\newcommand{\bd}{\begin{displaymath}}
\newcommand{\ed}{\end{displaymath}}
\newcommand{\mat}[4]{\left(\begin{array}{cc}{#1}&{#2}\\{#3}&{#4}
\end{array}\right)}
\newcommand{\matr}[9]{\left(\begin{array}{ccc}{#1}&{#2}&{#3}\\
{#4}&{#5}&{#6}\\{#7}&{#8}&{#9}\end{array}\right)}
\newcommand{\matrr}[6]{\left(\begin{array}{cc}{#1}&{#2}\\
{#3}&{#4}\\{#5}&{#6}\end{array}\right)}
\newcommand{\cvb}[3]{#1^{#2}_{#3}}
\newcommand\lsim{\raise0.3ex\hbox{$\;<$\kern-0.75em\raise-1.1ex
e\hbox{$\sim\;$}}}
\newcommand\gsim{\raise0.3ex\hbox{$\;>$\kern-0.75em\raise-1.1ex
\hbox{$\sim\;$}}}
\def\abs#1{\left| #1\right|}
\newcommand\simlt{\mathrel{\lower2.5pt\vbox{\lineskip=0pt\baselineskip=0pt
           \hbox{$<$}\hbox{$\sim$}}}}
\newcommand\simgt{\mathrel{\lower2.5pt\vbox{\lineskip=0pt\baselineskip=0pt
           \hbox{$>$}\hbox{$\sim$}}}}
\newcommand\unity{{\hbox{1\kern-.8mm l}}}
\newcommand{\eps}{\varepsilon}
\newcommand\ep{\epsilon}
\newcommand\ga{\gamma}
\newcommand\Ga{\Gamma}
\newcommand\om{\omega}
\newcommand\omp{{\omega^\prime}}
\newcommand\Om{\Omega}
\newcommand\la{\lambda}
\newcommand\La{\Lambda}
\newcommand\al{\alpha}
\newcommand{\ov}{\overline}
\renewcommand{\to}{\rightarrow}
\renewcommand{\vec}[1]{\mathbf{#1}}
\newcommand{\vect}[1]{\mbox{\boldmath$#1$}}
\newcommand\tm{{\widetilde{m}}}
\newcommand\mcirc{{\stackrel{o}{m}}}
\newcommand{\Dm}{\Delta m}
\newcommand{\dm}{\varepsilon}
\newcommand{\tanb}{\tan\beta}
\newcommand{\nbar}{\tilde{n}}
\newcommand\PM[1]{\begin{pmatrix}#1\end{pmatrix}}
\newcommand{\up}{\uparrow}
\newcommand{\down}{\downarrow}
\newcommand\omE{\omega_{\rm Ter}}
%

\newcommand{\Dsusy}{{susy \hspace{-9.4pt} \slash}\;}
\newcommand{\DCP}{{CP \hspace{-7.4pt} \slash}\;}
\newcommand{\mc}{\mathcal}
\newcommand{\gr}{\mathbf}
\renewcommand{\to}{\rightarrow}
\newcommand{\gtc}{\mathfrak}
\newcommand{\wh}{\widehat}
\newcommand{\br}{\langle}
\newcommand{\kt}{\rangle}


\def\lsim{\mathrel{\mathop  {\hbox{\lower0.5ex\hbox{$\sim$}
\kern-0.8em\lower-0.7ex\hbox{$<$}}}}}
\def\gsim{\mathrel{\mathop  {\hbox{\lower0.5ex\hbox{$\sim$}
\kern-0.8em\lower-0.7ex\hbox{$>$}}}}}

\newcommand\de{\partial}
\newcommand\brf{{\mathbf f}}
\newcommand\bbf{\bar{\bf f}}
\newcommand\bF{{\bf F}}
\newcommand\bbF{\bar{\bf F}}
\newcommand\bA{{\mathbf A}}
\newcommand\bB{{\mathbf B}}
\newcommand\bG{{\mathbf G}}
\newcommand\bI{{\mathbf I}}
\newcommand\bM{{\mathbf M}}
\newcommand\bY{{\mathbf Y}}
\newcommand\bX{{\mathbf X}}
\newcommand\bS{{\mathbf S}}
\newcommand\bb{{\mathbf b}}
\newcommand\bh{{\mathbf h}}
\newcommand\bg{{\mathbf g}}
\newcommand\bla{{\mathbf \la}}
\newcommand\bmu{\mathbf m }
\newcommand\by{{\mathbf y}}
\newcommand\bsig{\mbox{\boldmath $\sigma$} }
\newcommand\bunity{{\mathbf 1}}
\newcommand\cA{\mathcal{A}}
\newcommand\cB{\mathcal{B}}
\newcommand\cC{\mathcal{C}}
\newcommand\cD{\mathcal{D}}
\newcommand\cF{\mathcal{F}}
\newcommand\cG{\mathcal{G}}
\newcommand\cH{\mathcal{H}}
\newcommand\cI{\mathcal{I}}
\newcommand\cN{\mathcal{N}}
\newcommand\cM{\mathcal{M}}
\newcommand\cO{\mathcal{O}}
\newcommand\cR{\mathcal{R}}
\newcommand\cS{\mathcal{S}}
\newcommand\cT{\mathcal{T}}
\newcommand\eV{\mathrm{eV}}

\title{New Non-commutative and Higher Derivatives Quantum Mechanics from GUPs}

\author{Homa Shababi}
\email{h.shababi@scu.edu.cn}
\affiliation{Center for Theoretical Physics, College of Physical Science and Technology, Sichuan University, Chengdu 610065, P. R. China}
\author{Andrea Addazi}
\email{andrea.addazi@lngs.infn.it}
\affiliation{Center for Theoretical Physics, College of Physical Science and Technology, Sichuan University, Chengdu 610065, P. R. China}
\affiliation{INFN sezione Roma {\it Tor Vergata}, I-00133 Rome, Italy}

\date{\today}

\vspace{1cm}
\begin{abstract}

We explore a new class of Non-linear GUPs (NLGUP) showing the emergence of
a new non-commutative and higher derivatives quantum mechanics.
Within it, we introduce the shortest fundamental scale
as a UV fixed point in the NLGUP commutators $[X,P]=i\hbar f(P)$,
having in mind a fundamental highest energy threshold related to the Planck scale.
We show that this leads to lose commutativity of space coordinates,
that start to be dependent by the angular momenta of the system.
On the other hand, non-linear GUP must lead to a redefinition of the Schr\"odinger
equation to a new non-local integral-differential equation. We
also discuss the modification of the Dyson series in time-dependent perturbative approaches.
This may suggest that, in NLGUPs, non-commutativity and higher derivatives may be intimately interconnected
within a unified and coherent algebra.
We also show that Dirac and Klein-Gordon equations are extended with higher space-derivatives
according to the NLGUP. We compute momenta-dependent corrections to the dispersion velocity,
showing that the Lorentz invariance is deformed. We comment on possible implications in tests of light dispersion relations from Gamma-Ray-Bursts
or Blazars, with potential interests for future experiments such as LHAASO, HAWC and CTA.

\end{abstract}

\maketitle

\section{Introduction}
The Heisenberg Uncertainty Principle (HUP) is the essence of quantum mechanics.
 Is it possible to extend it in a Generalized form, i.e. a Generalized Uncertainty Principle (GUP),
compatible with Bohr correspondence principle?
This issue is also interesting from the point of view of quantum gravity, since we know that HUP is one of the main obstacle towards a consistent quantization of the Einstein geometrodynamical field.
A first simple proposal was to add quadratic powers as a simple non-linear extension of it, i.e. $[X_{i},P_{j}]=i\hbar (1+\beta |P|^{2}+...)$ \cite{KM},
which is also suggested as first effective corrections from scattering amplitudes in string theory \cite{ACV1,ACV2,ACV3}.
Neverthless, one would imagine that a full summation on all over perturbative and non-perturbative quantum gravity effects
may lead to a non-linear GUP (NLGUP) beyond quadratic corrections.

\vspace{0.1cm}

 In this paper, we wish to explore a class of GUP having, as a peculiarity, a UV momentum pole,
as $[X_{i},P_{j}]=i\hbar f(\beta P^{2})$ with $f\rightarrow \infty$ for $|P|\rightarrow 1/\sqrt{\beta}$.
We are particularly attracted by these models since they would imply, that at a critical UV momenta scale,
one would completely lose resolution on the $\Delta X \Delta P$ quantities and, therefore, on the total angular momenta.
This would imply that we may not probe length scales smaller than a critical length since fluctuations wildly diverge,
unitarizing cross-sections with the highest UV energy below the Planck scale.
The core of this proposal is to marry quantum uncertainties with Doubly Special Relativity \cite{55,56,57,58,59,60,P1,P2,P3}.
Within this framework, the presence of a UV fixed point in the NLGUP commutator of X and P
induces a fundamental shortest scale inside the quantum mechanical structure.
In our paper, we will explicitly show that, within such a new algebra extension, 
commutativity in the space coordinates is lost close to the critical energy $1/\sqrt{\beta}$.
We will also show that NLGUP implies new higher derivative terms appearing in the
wave function equation.

\vspace{0.3cm}

\section{Non-linear GUPs}

Let us start considering the following class of GUP models, that we may dub a-s-Non-linear-GUPs (asNLGUP).
\begin{eqnarray}\label{eq1}
[ X_i, P_j] =
\frac{ i\hbar  \delta_{ij} } { (1 - (\beta P^2)^{a})^{s} }\,\,,~~~
[ P_i, P_j]=0\,\,, ~~
[ X_i , X_j ] =\frac{ 2i\hbar  \beta  } { (1 - (\beta P^2)^{a})^{2s} }( P_i X_j - P_j X_i),
\end{eqnarray}
where $\beta$ is related to the GUP critical energy scale $\Lambda$ as $\beta=\Lambda^{-2}$,
$X,P$ are position and momenta operators, $a,s$ are two free parameters.
One should easily check that this class of model closes a self-consistence algebra set
while having an UV divergence for $|P|=\beta^{-1/2}$, having in mind that $\sqrt{\beta}$ may be related to the Planck length $l_{Pl}$.
This algebra is consistent with a redefinition of the standard position and momenta operator as follows:
\begin{equation}
\label{standard1}
P_i\equiv P_i, \,\,\,\,\, X_j\rightarrow X_{j}\frac{1}{(1 - (\beta P^2)^{a})^{s}}\, .
\end{equation}
Another important remark is that P-X would be conjugate variables in a generalized sense,
considering a measure factor to be included in the Fourier transforms.

The infinitesimal translation operator has a new non-linear form
\begin{equation}
\label{T}
T=1-i\frac{P \cdot dX}{(1-(\beta P^{2})^{a})^{s}}+O(dX^2)\rightarrow P =-i\hbar (1-(-\hbar^{2} \beta \nabla^{2})^{a})^{s} \nabla\, ,
\end{equation}
where $\nabla$ is with respect to standard $X$ variables of Q.M. Therefore, definitions in Eq. (\ref{standard1}) and Eq. (\ref{T}) are compatible each others.

This also suggests that angular momenta operators lose of any certainties around the UV fixed energy.
Indeed, Eq. (\ref{eq1}) implies a deformation of the standard angular momenta algebra (see Appendix A):
\begin{eqnarray}\label{eq5}
[ L_i, L_j]&=& \frac{i\hbar}{(1-(\beta P^2)^{a})^{s}}\left(X_i P_j-X_j P_i\right)= \frac{i\hbar}{(1-(\beta P^2)^{a})^{s}}\epsilon_{ijk}L_{k}\,\, ,
\end{eqnarray}
in compatibility with a redefinition of the angular momenta:
\begin{eqnarray}\label{new}
L_{i}=\frac{1}{(1-(\beta P^2)^{a})^{s}}\epsilon_{ijk}r_{j}p_{k}\, .
\end{eqnarray}
Substituting Eq. (\ref{new}) into Eq. (\ref{eq1}),
we obtain (see Appendix A)
\begin{eqnarray}\label{eq7}
[X_i,X_j]&=&\frac{ -2i\hbar  \beta } {(1 - (\beta P^2)^{a})^{s}}L_{ij}.
\end{eqnarray}
This implies a first unexpected aspect related to asNLGUP: in order to have a self-consistent algebra,
the space-coordinates have not to commute, and their non-commutativity depends from the angular momenta operator.
This would suggest a series of interesting facts.
One would imagine that non-commutativity of space coordinates depends on the angular momenta state of a
certain particle and, therefore, from angular momenta measures.
A measure of the angular moment on z-axis would
induce a non-commutativity of X and Y coordinates,
a measure on x-axis would induce it on Z,Y axis and so on.

In the special case $a=s=1$, it is worth to note that Eq. (\ref{new}) can be rewritten in the following form
\begin{eqnarray}\label{eq6-1}
L_{ij}=\left(X_i P_j-X_j P_i\right)\left(1-\beta P^2\right)^{-1} = \left(X_i P_j-X_j P_i\right)\left(1+\beta \hbar^2 \nabla^2\right)^{-1}\nonumber\\
= \left(X_i P_j-X_j P_i\right)\Big\{1-\beta \hbar^2 \nabla^2+\beta^2 \hbar^4 \nabla^4\nonumber\\ +...+\frac{-1(-2)(-3)...(-1-(n-1))}{n!}(\beta \hbar^2 \nabla^2)^n\Big\}.
\end{eqnarray}

This means that Eq. (\ref{eq7}) is not only non-commutative but also non-local in space-coordinates.
It is a remarkable feature of this theory that, if, for example, we imagine to measure X and later Y
this does not commute with the Y-X measure sequence and they are related each others through
a non-local derivative operator.

Now, it is worth to note that the whole deformations of standard quantum mechanics introduced
above lead to new extended Schr\"odinger equation.
In standard quantum mechanics, $H=i\hbar \frac{d}{dt}$ dictates the time evolution dynamics.
Here the Unitary operator has an infinitesimal structure that is deformed to a non-linear
functional:
\begin{equation}
\label{Unitary}
U=1-i\frac{H dt}{(1-(\beta P^{2})^{a})^{s}}+O(dt^2)\rightarrow H =i\hbar (1-(-\hbar^{2} \beta \nabla^{2})^{a})^{s} \frac{d}{dt}\, .
\end{equation}
This does not violate unitarity, at the fixed point:
the evolution operator $U$ for a finite time $t$ is
\begin{equation}
\label{Unitary}
U=e^{-\frac{iHt}{\hbar(1-(\beta P^2)^{a})^{s}}} \rightarrow U^{\dagger}U=UU^{\dagger}=1\, ,
\end{equation}
having unitarity automatically guaranteed.
From this definition we can also arrive to the extended Dyson series for a time-dependent perturbative approach
shown in our Appendix B.

The Schr\"odinger equation is extended to a new non-local differential equation
compatible with Eq. (\ref{standard1}) and Eq. (\ref{Unitary}):
\begin{equation}
\label{New0}
H\Psi=\Big[\frac{P^{2}}{2m}+V\{X/(1-(-\hbar^{2} \beta \nabla^2)^{a})^{s}\}\Big]\Psi,
\end{equation}
\begin{equation}
\label{New}
\rightarrow i \hbar \frac{d\Psi}{dt}=\frac{1}{(1-(-\hbar^{2} \beta \nabla^{2})^{a})^{s}}\Big[-(1-(-\hbar^{2} \beta \nabla^{2})^{a})^{2s}\frac{\hbar^{2} \nabla^{2}}{2m}+V\{X/(1-(-\hbar^{2} \beta \nabla^2)^{a})^s\}\Big]\, ,
\end{equation}
$$\rightarrow i \hbar \frac{d\Psi}{dt}= -(1-(-\hbar^{2} \beta \nabla^{2})^{a})^{s}\frac{\hbar^{2} \nabla^{2}}{2m}\Psi+\frac{1}{(1-(-\hbar^{2} \beta \nabla^{2})^{a})^{s}}V\{X/(1-(-\hbar^{2} \beta \nabla^2)^{a})^s\}\Psi\, .$$

In the case $\Psi(x,t)=\psi(x)e^{-iE t/\hbar}$, the Schr\"odinger equation would become a non-local time-independent
one:
\begin{equation}
\label{New2}
(1-(-\hbar^{2} \beta \nabla^{2})^{a})^{s}\frac{\hbar^{2} \nabla^{2}\psi(x)}{2m}+\frac{1}{(1-(-\hbar^{2} \beta \nabla^{2})^{a})^{s}}V\{X/(1-(-\hbar^{2} \beta \nabla^2)^{a})^s\}\psi(x)=E \psi(x)\, .
\end{equation}

Eq. (\ref{standard1}) implies that the kinetic term of the Hamiltonian
$H_{0}=P^2/2m$ is untouched.
On the other hand, interaction potentials are deformed
from their dependence by $X$ operators.
It is worth to remind that $\nabla=\nabla_{X_{B}}\neq \nabla_{X}$, as stated above. 

A generic central potential
$V=\alpha r^{\beta}$
would be deformed as
$V=\alpha r^{\beta}/(1-(\beta P^{2})^{m})^{k\beta}$.
This can be reinterpreted as a new energy dependent
re-normalized coupling
\begin{equation}\label{alphaha}
\alpha_{R}=\alpha (1-(\beta P^{2})^{a})^{-s\beta}\, .
\end{equation}
Conversely, in the case of the electric or the gravitational field,
we would have $\beta=-1$ and therefore
\begin{equation}\label{alphaha}
\alpha_{R}=\alpha (1-(\beta P^{2})^{a})^{s}\, .
\end{equation}
This would suggest  that the e.m and gravitational couplings would eventually flow
to a U.V. fixed point when they flow to zero and ${\rm lim}_{P\rightarrow \beta^{-1/2}}\alpha_{R}\rightarrow 0$.



Let us consider the deformation of the Klein-Gordon equation in the relativistic
regime $v\simeq c$. In NLGUP considered above this would read as
\begin{equation}
\label{KG}
P^{2}+m^2 =E^2 \rightarrow (1-(-\hbar^{2}\nabla^{2})^a)^{2s} \Box \Psi+m^{2} \Psi= 0\,\,\,\, (c=1)\, .
\end{equation}

From Klein-Gordon equation, we can obtain the Schr\"odinger equation
by performing the non-relativistic limit as follows:
\begin{equation}
\label{Perf}
\Psi=\psi e^{-im_{0}t/\hbar(1-(\beta P^{2})^{a})^{s}},\,\,\, (c=1)\, ,
\end{equation}
\begin{equation}
\label{v}
v<<1 \rightarrow |\dot{\psi}|<<1\, .
\end{equation}

Within Eqs. (\ref{Perf}) and (\ref{v}) assumptions,
we obtain
\begin{equation}
\label{dotpsipsi}
\dot{\Psi}=\Big(-\frac{im_{0}}{\hbar(1-(\beta P^{2})^{a})^{s}}  \Big)\psi e^{-im_{0}t/\hbar(1-(\beta P^{2})^{a})^{s}}\, ,
\end{equation}
\begin{equation}
\label{jaa}
\ddot{\Psi}=\Big(-\frac{2im_{0}}{\hbar(1-(\beta P^{2})^{a})^{s}}\dot{\psi}+\frac{m_{0}^{2}}{\hbar^{2}(1-(\beta P^{2})^{a})^{2s}}\psi\Big)e^{-im_{0}t/\hbar(1-(\beta P^{2})^{a})^{s}}\, ,
\end{equation}
and inserting Eq. (\ref{jaa}) into the modified Klein-Gordon equation, we obtain
\begin{equation}
\label{kk}
(1-(\beta P^{2})^{a})^{2s}\Big(-\frac{2im_{0}}{\hbar(1-(\beta P^{2})^{a})^{s}}\dot{\psi}+\frac{m_{0}^{2}}{\hbar^{2}(1-(\beta P^{2})^{a})^{2s}}\psi\Big)-(1-(\beta P^{2})^{a})^{2s}\nabla^2\psi -m_{0}^{2}\psi=0\, ,
\end{equation}
and it is easy to see that mass terms $m_{0}^{2}$ cancel each others and we obtain the modified Schr\"odinger equation for a free-particle $(V=0)$ in Eq. (\ref{New2}).

It is also easy to extended the Dirac equation, from the modified definition of energy and momenta as
\begin{equation}
\label{Diract}
i(1-(-\hbar^{2}\nabla^{2})^{a})^{s}\gamma_{\mu}\partial^{\mu}\Psi+m^{2}\Psi=0\, ,
\end{equation}
which is compatible with Schr\"odinger equation in the non-relativistic limit.

Finally, we can also extend the electromagnetic and gravitational waves equations as
\begin{equation}
\label{ele}
(1-(-\hbar^{2}\nabla^{2})^a)^{2s} \Box A_{\mu}=0,\,\,\,\,\,\,\,\,\, (1-(-\hbar^{2}\nabla^{2})^a)^{2s} \Box h_{\mu\nu}=0.
\end{equation}

From the extended wave equations,
the standard dispersion relations are modified as an effect of Lorentz invariance deformation.
Indeed, considering a standard wave solution $\Psi=A\, e^{-iE_{B} t+ip_{B}x}$
and inserting it inside the K.G. equation,
we obtain, for $a=1$ and $s=1$ and zero mass,
\begin{equation}
\label{GUPdispersion}
E_{B}^{2}=c^{2}p^{2}_{B}+\beta p^{4}_{B},\,\,\, v^{2}=c^{2}(1+ \beta \, p^{2}_{B})\, .
\end{equation}

For a general NLGUP modification, we find the first correction as
\begin{equation}
\label{GUPdispersion}
v=c(1+ (s/2)(\beta \, p^{2}_{B})^{a})+O(p^{2}_{B})^{a+1}\, .
\end{equation}
Indeed, we do not see any obstruction in having arbitrary small rational number $a$, except experimental constrains.
If $a=1/2$, the UV singularity in Eq.1 would correspond to a string in the complex Energy plane rather then a single point,
which would smell as introducing a new non-locality in scattering amplitudes.

\section{Conclusion and Remarks}

In this paper, we discussed a new class of non-linear GUP as an attempt to
implement a minimal length in quantum mechanics by extending the standard Heisenberg's uncertainty relation.
We shown that this class of NLGUP implies that space-coordinates are non-commutative;
the commutators are dependent by the angular momenta operators.
This is a new feature that, so far as we know, was never {\it met} in any other theories of non-commutative space-time.
Then, we found that NLGUP implies an extension of the Schr\"odinger, Klein-Gordon and Dirac equations
including new higher spatial derivative terms.
Such a result reminds the Ho\v{r}ava-Lishfitz theory \cite{Horava:2009uw}, 
as a higher space-derivatives extension of the lagrangian, without introducing any new time derivatives, i.e. without introducing any ghosts.

Another direct consequence is the modification of dispersion relations in {\it vacuo} with
momentum-dependent corrections to wave and particle speeds.
Surely, this is a manifestation of a deformation of the Lorentz algebra.
Therefore, modified dispersion relations open intriguing phenomenological channels for
this model, as a test of
quantum mechanics foundations and quantum gravity.
Indeed tests of dispersion velocities of high energy gamma-rays from Gamma-Ray-Bursts or Blazars
were proposed by many authors as a possible new frontier of quantum gravity phenomenology
\cite{Fairbairn:2014kda,Ellis:2018lca,Ellis:2018ogq,AmelinoCamelia:2008qg,Jacobson:2004rj,Vasileiou:2015wja,Xu:2016zxi,Zhang:2018otj,Xu:2018ien}.
 Therefore, Non-linear GUP strongly motivates tests of Lorentz deformations
from future Very High Energy Gamma-Rays detectors such as LHAASO \cite{LHAASO}, CTA \cite{CTA} and HAWC \cite{HAWC}.
In the case $s=1$ and $a=1/2$, Eq.\ref{GUPdispersion} can be constrained up to $1/\sqrt{\beta}\geq 10^{16}\, {\rm GeV}$ from current high energy neutrinos in IceCube \cite{Ellis:2018ogq}.

Finally, we suspect that NLGUP can deform the Spin-Statistics of standard quantum mechanics,
with possible important implications in searches of Pauli Exclusion Principle Violations from Quantum gravity in underground experiments
\cite{A1,A2,A3,A4}.

\section*{Appendix A}

Here, we show a more detailed proof of self-consistency of Non-linear GUPs with the deformation of the angular momentum algebra
as well as non-commutativity of space-coordinates. For formal simplicity, we will just consider the case $a=s=1$ as generalizations of it are easily understood.
First, let us express the angular momenta algebra just in terms of position and momenta operators:
\begin{eqnarray}\label{eq2}
[ L_i, L_j]&=&\left[\epsilon_{ik\ell}X_k P_\ell,\epsilon_{jmn}X_m P_n\right]=\epsilon_{ik\ell}\epsilon_{jmn}[X_k P_\ell,X_m P_n]
\nonumber\\&=&\epsilon_{ik\ell}\epsilon_{jmn}\Big\{[X_k,X_m]P_\ell P_n +X_m [X_k,P_n] P_\ell +X_k[P_\ell,X_m]P_n+X_k X_m[P_\ell,P_n]\Big\}.
\end{eqnarray}
Now, using Eq. (\ref{eq1}), we obtain:
\begin{eqnarray}\label{eq3}
[ L_i, L_j]&=&\epsilon_{ik\ell}\epsilon_{jmn}\left\{\frac{ 2i\hbar \beta}{(1-\beta P^2)^2}( P_k X_m - P_m X_k)P_\ell P_n+X_m \frac{i\hbar \delta_{kn}}{1-\beta P^2}P_\ell+X_k \frac{-i\hbar \delta_{\ell m}}{1-\beta P^2}P_n \right\}\nonumber\\&=&\frac{ 2i\hbar \beta}{(1-\beta P^2)^2}\Big\{\epsilon_{ik\ell}\epsilon_{jmn}P_k X_m P_\ell P_n-\epsilon_{ik\ell}\epsilon_{jmn}P_m X_k P_\ell P_n \Big\}\nonumber\\&&+\frac{i\hbar}{1 -\beta P^2}\Big\{\epsilon_{ik\ell}\epsilon_{jmn}\delta_{kn}X_m P_\ell -  \epsilon_{ik\ell}\epsilon_{jmn}\delta_{\ell m}X_k P_n\Big\}\nonumber\\&=&\frac{ 2i\hbar \beta}{(1-\beta P^2)^2}\left\{\epsilon_{ik\ell}\epsilon_{jmn}P_k \left(P_\ell X_m + \frac{i\hbar\delta_{\ell m}}{1-\beta P^2}\right)P_n - \epsilon_{ik\ell}\epsilon_{jmn}P_m \left(P_n X_k + \frac{i\hbar\delta_{k n}}{1-\beta P^2}\right)P_\ell\right\}\nonumber\\&&+\frac{i\hbar}{1 -\beta P^2}\Big\{-\epsilon_{i\ell k}\epsilon_{jmn}X_m P_\ell+\epsilon_{ik\ell}\epsilon_{jn\ell} X_k P_n\Big\}\nonumber\\&=&\frac{ 2i\hbar \beta}{(1-\beta P^2)^2}\left\{\epsilon_{ik\ell}\epsilon_{jmn}P_k P_\ell X_m P_n+\frac{i\hbar \epsilon_{ik\ell}\epsilon_{jmn}\delta_{\ell m}P_k P_n}{1-\beta P^2}-\epsilon_{ik\ell}\epsilon_{jmn}P_m P_n X_k P_\ell-\frac{i\hbar \epsilon_{ik\ell}\epsilon_{jmn}\delta_{kn}P_m P_\ell}{1-\beta P^2}\right\}\nonumber\\&&+\frac{i\hbar}{1 -\beta P^2}\Big\{-\left(\delta_{ij}\delta_{\ell m}-\delta_{im}\delta_{j\ell}\right)X_m P_\ell+ \left(\delta_{ij}\delta_{kn}-\delta_{in}\delta_{jk}\right)X_k P_n \Big\},
\end{eqnarray}
where the first and third sentences of the first bracket are equal to zero, due to the product of symmetric and anti-symmetric arrays.
Then, we can simplify if as
\begin{eqnarray}\label{eq4}
[ L_i, L_j]&=&\frac{ 2i\hbar \beta}{(1-\beta P^2)^2}\left\{\frac{i\hbar \epsilon_{ik\ell}\epsilon_{jmn}\delta_{\ell m}P_k P_n}{1-\beta P^2}-\frac{i\hbar \epsilon_{ik\ell}\epsilon_{jmn}\delta_{kn}P_m P_\ell}{1-\beta P^2}\right\}\nonumber\\&&+\frac{i\hbar}{1 -\beta P^2}\Big\{-\delta_{ij}\delta_{\ell m}X_m P_\ell+ \delta_{im}\delta_{j\ell}X_m P_\ell+\delta_{ij}\delta_{kn}X_k P_n-\delta_{in}\delta_{jk}X_k P_n\Big\}\nonumber\\&=&\frac{ -2\hbar^{2} \beta}{(1-\beta P^2)^3}\Big\{-\epsilon_{ik\ell}\epsilon_{jn\ell}P_k P_n+\epsilon_{i\ell k}\epsilon_{jmk}P_m P_\ell\Big\}+\frac{i\hbar}{1 -\beta P^2}\Big\{-\delta_{ij}X_\ell P_\ell+X_i P_j+-\delta_{ij}X_k P_k-X_i P_j\Big\}\nonumber\\&=& \frac{ -2\hbar^{2} \beta}{(1-\beta P^2)^3}\Big\{-\left(\delta_{ij}\delta_{kn}-\delta_{in}\delta_{kj}\right)P_k P_n+ \left(\delta_{ij}\delta_{\ell m}-\delta_{im}\delta_{j\ell}\right)P_m P_\ell \Big\}+\frac{i\hbar}{1-\beta P^2}\left(X_i P_j-X_j P_i\right)\nonumber\\&=& \frac{ -2\hbar^{2} \beta}{(1-\beta P^2)^3}\Big\{ -\delta_{ij}P_k P_k +P_j P_i+\delta_{ij}P_\ell P_\ell-P_i P_i\Big\}+\frac{i\hbar}{1-\beta P^2}\left(X_i P_j-X_j P_i\right),
\end{eqnarray}
Finally, Eq. (\ref{eq4}) will lead to
\begin{eqnarray}\label{eq5}
[ L_i, L_j]&=& \frac{i\hbar}{1-\beta P^2}\left(X_i P_j-X_j P_i\right).
\end{eqnarray}

\section*{Appendix B: Extended Dyson series}

The easiest way to derive the time-ordered perturbation expansion is to
use the S-operator in the following form
\begin{eqnarray}\label{eq32}
S=U(\infty,-\infty),
\end{eqnarray}
where with the Hamiltonian in the form of $H=H_0+V$ and in the presence of GUP (\ref{eq1})
\begin{eqnarray}\label{eq33}
U(\tau,\tau_0)= \exp\left(\frac{iH_{0}\tau}{\hbar\left(1-(-\hbar^2 \beta\nabla^2)^a\right)^s}\right)\exp\left(\frac{iH(\tau-\tau_0)}{\hbar\left(1-(-\hbar^2 \beta\nabla^2)^a\right)^s}\right)\exp\left(\frac{-iH_{0}\tau_0}{\hbar\left(1-(-\hbar^2 \beta\nabla^2)^a\right)^s}\right).
\end{eqnarray}
Now, differentiating Eq. (\ref{eq33}) with respect to $\tau$ gives the following
differential equation
\begin{eqnarray}\label{eq34}
i\hbar\left(1-(-\hbar^2 \beta\nabla^2)^a\right)^s\frac{\partial U(\tau,\tau_0)}{\partial \tau}= V(\tau)U(\tau,\tau_0),
\end{eqnarray}
where
\begin{eqnarray}\label{eq35}
V(t)=\exp\left(\frac{iH_{0}t}{\hbar\left(1-(-\hbar^2 \beta\nabla^2)^a\right)^s}\right) V \exp\left(\frac{-iH_{0}t}{\hbar\left(1-(-\hbar^2 \beta\nabla^2)^a\right)^s}\right).
\end{eqnarray}
Now, Eq. (\ref{eq34}) as well as the initial condition
$U(\tau_0,\tau_0) = 1$ is obviously satisfied by the solution of the integral equation as
\begin{eqnarray}\label{eq36}
U(\tau,\tau_0) = 1- \frac{i}{\hbar\left(1-(-\hbar^2 \beta\nabla^2)^a\right)^s}\int_{\tau_0}^{\tau}dt V(t)U(t,\tau_0).
\end{eqnarray}
Iterating this integral equation, we obtain an expansion for $U(\tau,\tau_0)$
in powers of $V$
\begin{eqnarray}\label{eq37}
U(\tau,\tau_0) &=& 1- \frac{i}{\hbar\left(1-(-\hbar^2 \beta\nabla^2)^a\right)^s}\int_{\tau_0}^{\tau}dt_1 V(t_1)+\frac{(-i)^2}{\hbar^2\left(1-(-\hbar^2 \beta\nabla^2)^a\right)^{2s}}\int_{\tau_0}^{\tau}dt_1 \int_{\tau_0}^{t_1}dt_2 V(t_1)V(t_2)\nonumber\\&&+\frac{(-i)^3}{\hbar^3\left(1-(-\hbar^2 \beta\nabla^2)^a\right)^{3s}}\int_{\tau_0}^{\tau}dt_1 \int_{\tau_0}^{t_1}dt_2 \int_{\tau_0}^{t_2}dt_3V(t_1)V(t_2)V(t_3)+... .
\end{eqnarray}
Now, if we set $\tau=\infty$ and $\tau_0 = -\infty$, the perturbation expansion for the
S-operator obtains as
\begin{eqnarray}\label{eq38}
S&=& 1- \frac{i}{\hbar\left(1-(-\hbar^2 \beta\nabla^2)^a\right)^s}\int_{-\infty}^{\infty}dt_1 V(t_1)+\frac{(-i)^2}{\hbar^2\left(1-(-\hbar^2 \beta\nabla^2)^a\right)^{2s}}\int_{-\infty}^{\infty}dt_1 \int_{-\infty}^{t_1}dt_2 V(t_1)V(t_2)\nonumber\\&&+\frac{(-i)^3}{\hbar^3\left(1-(-\hbar^2 \beta\nabla^2)^a\right)^{3s}}\int_{-\infty}^{\infty}dt_1 \int_{-\infty}^{t_1}dt_2 \int_{-\infty}^{t_2}dt_3V(t_1)V(t_2)V(t_3)+... .
\end{eqnarray}
There is a way of rewriting Eq. (\ref{eq38}) that proves very useful in
carrying out manifestly Lorentz-invariant calculations. Define the time-
ordered product of any time-dependent operators as the product with
factors arranged so that the one with the latest time-argument is placed
leftmost, the next-latest next to the leftmost, and so on. For instance,
$$T\{V(t)\}=V(t),$$
$$T\{V(t_1),V(t_2)\}=\theta\left(t_1-t_2\right)V(t_1)V(t_2)+\theta\left(t_2-t_1\right)V(t_2)V(t_1),$$
and so on, where $\theta(\tau)$ is the step function, i.e.,
$$\theta(\tau) =
\begin{cases}
+1 ~~~  \tau > 0,\\
0 ~~~~~~ \tau < 0.
\end{cases}$$
The time-ordered product of $n$ $Vs$ is a sum over all $n!$
permutations of the $Vs$, each of which gives the same integral over all $t_1...t_n$. Hence, Eq. (\ref{eq38}) can be written in the following form
\begin{eqnarray}\label{eq39}
S&=& 1+\sum_{n=1}^{\infty}\frac{(-i)^n}{\hbar^n\left(1-(-\hbar^2 \beta\nabla^2)^a\right)^{sn} n!}\int_{-\infty}^{\infty}dt_1 dt_2...dt_n T\{V(t_1)V(t_2)...V(t_n)\},
\end{eqnarray}
which is called modified Dyson series.


\begin{thebibliography}{99}

\bibitem{KM} A. Kempf, G. Mangano, and R. B. Mann, Phys. Rev. D {\bf 52}, 1108 (1995).
\bibitem{ACV1} G. Veneziano, Europhys. Lett. {\bf 2}, 199 (1986).
\bibitem{ACV2} D. Amati, M. Ciafaloni and G. Veneziano, Phys. Lett. B {\bf 216}, 41 (1989).
\bibitem{ACV3} D. Amati, M. Ciafaloni and G. Veneziano, Phys. Lett. B {\bf 197}, 81 (1987).
\bibitem{55} A. F. Ali, S. Das and E. C. Vagenas, Phys. Lett. B {\bf 678}, 497 (2009).
\bibitem{56} S. Das, E. C. Vagenas and A. F. Ali, Phys. Lett. B {\bf 690}, 407 (2010).
\bibitem{57} A. F. Ali, S. Das, and E. C. Vagenas, Phys. Rev. D {\bf 84}, 044013 (2011).
\bibitem{58} J. Magueijo and L. Smolin, Phys. Rev. Lett. {\bf 88}, 190403 (2002).
\bibitem{59} J. Magueijo and L. Smolin, Phys. Rev. D {\bf 71}, 026010 (2005).
\bibitem{60} J. L. Cortes and J. Gamboa, Phys. Rev. D {\bf 71}, 065015 (2005).
\bibitem{P1} P. Pedram, Phys. Lett. B {\bf 714}, 317 (2012).
\bibitem{P2} P. Pedram, Phys. Lett. B {\bf 718}, 638 (2012).
\bibitem{P3} H. Shababi and W. S. Chung, Phys. Lett. B {\bf 70}, 445 (2017).
\bibitem{Horava:2009uw} P.~Ho\v{r}ava, Phys.\ Rev.\ D {\bf 79}, 084008 (2009).
\bibitem{Fairbairn:2014kda} M.~Fairbairn, A.~Nilsson, J.~Ellis, J.~Hinton and R.~White, JCAP {\bf 1406}, 005 (2014).
\bibitem{Ellis:2018lca} J.~Ellis, R.~Konoplich, N.~E.~Mavromatos, L.~Nguyen, A.~S.~Sakharov and E.~K.~Sarkisyan-Grinbaum, Phys.\ Rev.\ D {\bf 99}, 083009 no.8 (2019.
\bibitem{Ellis:2018ogq} J.~Ellis, N.~E.~Mavromatos, A.~S.~Sakharov and E.~K.~Sarkisyan-Grinbaum, Phys.\ Lett.\ B {\bf 789}, 352 (2019).
\bibitem{AmelinoCamelia:2008qg} G.~Amelino-Camelia, Living Rev.\, Rel.\,  {\bf 16}, 5 (2013).
\bibitem{Jacobson:2004rj} T.~Jacobson, S.~Liberati and D.~Mattingly, Lect.\ Notes Phys.\  {\bf 669}, 101 (2005).
\bibitem{Vasileiou:2015wja} V.~Vasileiou, J.~Granot, T.~Piran and G.~Amelino-Camelia, Nature Phys.\  {\bf 11}, 344 no.4 (2015).
\bibitem{Xu:2016zxi} H.~Xu and B.~Q.~Ma, Astropart.\ Phys.\  {\bf 82}, 72 (2016).
\bibitem{Zhang:2018otj} X.~Zhang and B.~Q.~Ma, Phys.\ Rev.\ D {\bf 99}, 043013 no.4 (2019).
\bibitem{Xu:2018ien} H.~Xu and B.~Q.~Ma, JCAP {\bf 1801}, 050 (2018).
\bibitem{LHAASO} X.~Bai {\it et al.}, arXiv:1905.02773 [astro-ph.HE].
\bibitem{CTA} B.~S.~Acharya {\it et al.} [CTA Consortium], Astropart.\ Phys.\  {\bf 43}, 3 (2013).
\bibitem{HAWC} A.~Albert {\it et al.} [HAWC Collaboration], arXiv:1911.08070 [astro-ph.HE].
\bibitem{A1} A. Addazi, P. Belli, R. Bernabei and A. Marciano, Chin. Phys. C {\bf 42}, 094001 no.9 (2018).
\bibitem{A2} A. Addazi and R. Bernabei, Mod. Phys. Lett. A {\bf 34}, 1950236 no.29  (2019).
\bibitem{A3} A. Addazi and A. Marciano, arXiv:1811.06425 [hep-ph], accepted in IJMPD.
\bibitem{A4} A. Addazi and R. Bernabei, arXiv:1901.00390 [hep-ph], accepted in IJMPD.

\end{thebibliography}
\end{document}